\documentclass[twocolumn]{aastex631dm}

\begin{document}

\title{Chromospheric modeling of the active M3V star G\,80-21 with RH1.5D}

\author[0000-0001-5193-1727]{Shuai Liu}
\affiliation{Key Laboratory of Optical Astronomy, 
National Astronomical Observatories, 
 Chinese Academy of Science, Beijing 100011, People's Republic of China}
\affiliation{Instituto de Astrof\'{i}sica de Canarias, V\'{i}a 
L\'{a}ctea, 38205 La Laguna, Tenerife, Spain}

\author[0000-0002-9681-6148]{Huigang Wei}
\affiliation{Key Laboratory of Optical Astronomy, 
National Astronomical Observatories, 
 Chinese Academy of Science, Beijing 100011, People's Republic of China}
\affiliation{School of Astronomy and Space Science, University of Chinese Academy of Sciences, Beijing 100049, China}

\author[0000-0002-0349-7839]{Jianrong Shi}
\affiliation{Key Laboratory of Optical Astronomy, 
National Astronomical Observatories, 
 Chinese Academy of Science, Beijing 100011, People's Republic of China}
\affiliation{School of Astronomy and Space Science, University of Chinese Academy of Sciences, Beijing 100049, China}
\affiliation{School of Physics and Technology, Nantong University, Nantong 226019, China}
\correspondingauthor{Jianrong Shi}
\email{sjr@nao.cas.cn}

\author[0000-0002-4569-1568]{Wenxian Li}
\affiliation{National Astronomical Observatoriess, 
 Chinese Academy of Science, Beijing 100011, People's Republic of China}
\affiliation{Key Laboratory of Solar Activity and Space Weather, National Space Science Center, Chinese Academy of Sciences, Beijing 100190, China}

\author[0000-0003-3474-5118]{Henggeng Han}
\affiliation{Key Laboratory of Optical Astronomy, 
National Astronomical Observatories, 
 Chinese Academy of Science, Beijing 100011, People's Republic of China}

\author[0000-0002-2874-2706]{Jifeng Liu}
\affiliation{Key Laboratory of Optical Astronomy, 
National Astronomical Observatories, 
 Chinese Academy of Science, Beijing 100011, People's Republic of China}
\affiliation{School of Astronomy and Space Science, University of Chinese Academy of Sciences, Beijing 100049, China}

\author[0000-0002-2967-4522]{Shangbin Yang}
\affiliation{School of Astronomy and Space Science, University of Chinese Academy of Sciences, Beijing 100049, China}
\affiliation{National Astronomical Observatoriess, 
 Chinese Academy of Science, Beijing 100011, People's Republic of China}




\begin{abstract}
This study investigates the active regions of the M3.0V star G\,80-21 using the observed data from the CARMENES project with synthetic spectra generated by the RH1.5D radiative transfer code. The CARMENES project aims to search for exoplanets around M dwarfs using high-resolution near-infrared and optical echelle spectrographs. By comparing the observed data and models for the chromospheric lines of H$_\alpha$ and the bluest \ion{Ca}{2} infrared triplet line, we obtain the best-fit models for this star. The optimal fitting for the observed spectrum of G\,80-21 is achieved by employing two active areas in conjunction with an inactive regions, with a calcium abundance of [Ca/H] = $-$0.4. This combination successfully fits all the observed data across varying ratios. The minor active component consistently comprises approximately 18\% of the total (ranging from 14\% to 20\%), which suggests that the minor active component is likely located in the polar regions. Meanwhile, the major active component occupies a variable proportion, ranging from 51\% to 82\%. Our method allows for the determination of the structure and size of stellar chromospheric active regions by analyzing high-resolution observed spectra.

\end{abstract}



\section{Introduction} \label{sec:intro}

M dwarfs, which are small and cool, have lower mass and surface temperatures than stars like the Sun. They are the most common type of stars in the Milky Way galaxy, accounting for about 70\% of the total number of stars \citep[e.g.][]{2010AJ....139.2679B,2016PhR...663....1S}. 
Due to their abundance and long-lived nature, M dwarfs are prime targets for studying stellar activity and searching for exoplanets. Their small size makes it easier to detect Earth-sized rocky planets in their habitable zones, as these planets can induce velocity changes in the star on the order of meters per second
\citep[e.g.][]{2007AsBio...7...85S,2009IAUS..253...37I,2015ApJ...807...45D,2016PhR...663....1S}.

The chromosphere is a crucial part of a star’s atmosphere, located between the photosphere and the corona, and it significantly influences the spectral characteristics of a star \citep{1981ApJ...250..709A,2000ssma.book.....S,2003rtsa.book.....R,2008LRSP....5....2H,2012LRSP....9....1R}. Studying chromospheric structures is essential for understanding stellar dynamics, as they are central to magnetic activity, influencing phenomena like starspots and flares \citep{2017ApJ...834...85N,1996AJ....112.2799H}.
For M dwarf stars, the chromosphere exhibits significant activity due to their deep convective zone, high rotation rates, and the interaction between convection and magnetic fields; all of them contribute to the star’s overall magnetic activity \citep{1986ApJS...61..531S,1998A&A...331..581D,2008ApJ...684.1390R,2017ApJ...834...85N,2021A&ARv..29....1K}. This layer is particularly notable for its strong emission in specific spectral lines \citep{1981ApJS...46..159W,2004AJ....128..426W}, with the H$_\alpha$ and Ca II H \& K lines being the most prominent \citep{1996AJ....112.2799H,2023A&A...672A..37I}. Recently, some researches \citep{2011MNRAS.414.2629M,2017A&A...605A.113M} indicated that the \ion{Ca}{2} infrared triplet (IRT) lines are as effective as the \ion{Ca}{2} H \& K lines in assessing chromospheric activity. 


Traditional one-dimensional models have achieved significant success but are limited in simulating chromospheric activity due to their inability to capture lateral structures and dynamic effects in stellar atmospheres. \citet{2011ApJ...736...69U} extensively discussed these limitations, pointing out that neglecting convective motions, nonlinearities in temperature and density, and nonlinear effects in computing molecular equilibrium can lead to inaccurate spectral interpretations.
RH1.5D\citep{2015A&A...574A...3P} is a modified version of the RH radiative transfer code\citep{1991A&A...245..171R,1992A&A...262..209R,2001ApJ...557..389U}, offering improved convenience for accurately diagnosing phenomena in simulations and better comparison with observations. This model addresses these issues by employing a hybrid approach that solves radiative transfer and hydrodynamic equations in one-dimensional space (vertically) while considering average effects laterally. This balance between computational complexity and physical accuracy allows RH1.5D to effectively simulate chromospheric emission lines by using high-precision methods for radiative transfer, accounting for non-local thermodynamic equilibrium (Non-LTE) effects \citep{1997ApJ...481..500C}, and including dynamic processes such as convection and waves \citep{2007A&A...473..625L}. Additionally, it incorporates magnetic field influences through magnetohydrodynamic (MHD) equations, simulating its impact on atmospheric structure and radiation \citep{2016ApJ...826..144S}.

Key features of RH1.5D include handling multi-level radiative transfer, considering partial frequency redistribution (PRD) effects, and incorporating Zeeman polarization due to magnetic fields. This multi-level approach accurately represents atomic and molecular transitions, which are essential for realistic spectral synthesis \citep{2012A&A...539A..39C}. PRD effect is crucial for modeling chromospheric emission lines where scattering processes affect the line profiles \citep{2013ApJ...772...89L}, and the capability to simulate the Zeeman effect allows for the study of magnetic field impacts on stellar spectra \citep{2012ASPC..456...59S}. 

In this study, we utilize the RH1.5D code to compute spectra from chromospheric model atmospheres with a parametrized temperature stratification for the M3.0-type star G\,80-21, and then compare the simulated spectra with high-resolution, high signal-to-noise ratio (SNR) spectra obtained with the CARMENES\footnote{Calar Alto high-Resolution search for M dwarfs with Exoearths with Near-infrared and optical \'{E}chelle Spectrographs.} spectrograph. 
Section \ref{sec:obs} describes these observations, while Section \ref{sec:mod}  outlines both the model construction process and the method of adjusting active-to-quiet region ratios in model spectra to match observed variations, using the inactive star LP 819-17 as a reference for the quiet region model.
Section \ref{sec:result} explores the impact of magnetic fields on chromospheric spectra and investigates combinations of active models to better fit the observed spectra of the sample star, highlighting varying contributions of active regions over time. Finally, in Section \ref{sec:con}, we present our conclusions.

\section{Observations} \label{sec:obs}
The CARMENES \citep{2018SPIE10702E..0WQ}, is mounted on the 3.5\,m telescope at the Calar Alto Observatory. This highly precise instrument operates its visual channel (VIS) within a wavelength range of 5200 to 9600\,\AA, while its infrared channel (NIR) spans from 9600 to 17100\,\AA. The VIS channel boasts a spectral resolution of approximately R$\sim$ 94,600, and the NIR channel has a resolution of R $\sim$ 80,400. Although the spectra from CARMENES do not include the \ion{Ca}{2} H \& K lines, they do cover the \ion{Ca}{2} IRT lines. Recent studies have shown that the \ion{Ca}{2} IRT lines can serve as excellent substitutes for the traditional blue \ion{Ca}{2} H \& K lines as chromospheric activity  indicators  \citep{2011MNRAS.414.2629M,2017A&A...605A.113M}. Additionally, the spectra cover other chromospheric activity lines, such as H$_\alpha$, \ion{Na}{1}D, and \ion{He}{1} D3 lines. For our analysis, we utilized the data from the VIS channel to study the activity in the H$_\alpha$ and \ion{Ca}{2} IRT lines.

In this study, an active M3.0V dwarf G\,80-21 is selected from the CARMENES Data Release~1 (2016-2020), which includes 19,633 spectra for 362 targets \citep{2023A&A...670A.139R}, providing valuable insights into M dwarf stars. 
The active M3.0V dwarf star G\,80-21, selected for this study, exhibits strong chromospheric activity, with prominent emission lines observed in both H$_\alpha$ and \ion{Ca}{2}, making it an ideal candidate for investigating chromospheric spectral line formation mechanisms. Observationally, this star underwent 11 time-domain observations between August 30, 2016, and January 12, 2017. The SNR for the H$_\alpha$ line ranged from 31 to 61, while the SNR for the \ion{Ca}{2} triplet lines varied from 51 to 96. The airmass during these observations was between 1.28 and 1.39, indicating consistent atmospheric conditions when the spectral data were acquired.

\begin{figure*}[htb]
\includegraphics[width=1.0\hsize]{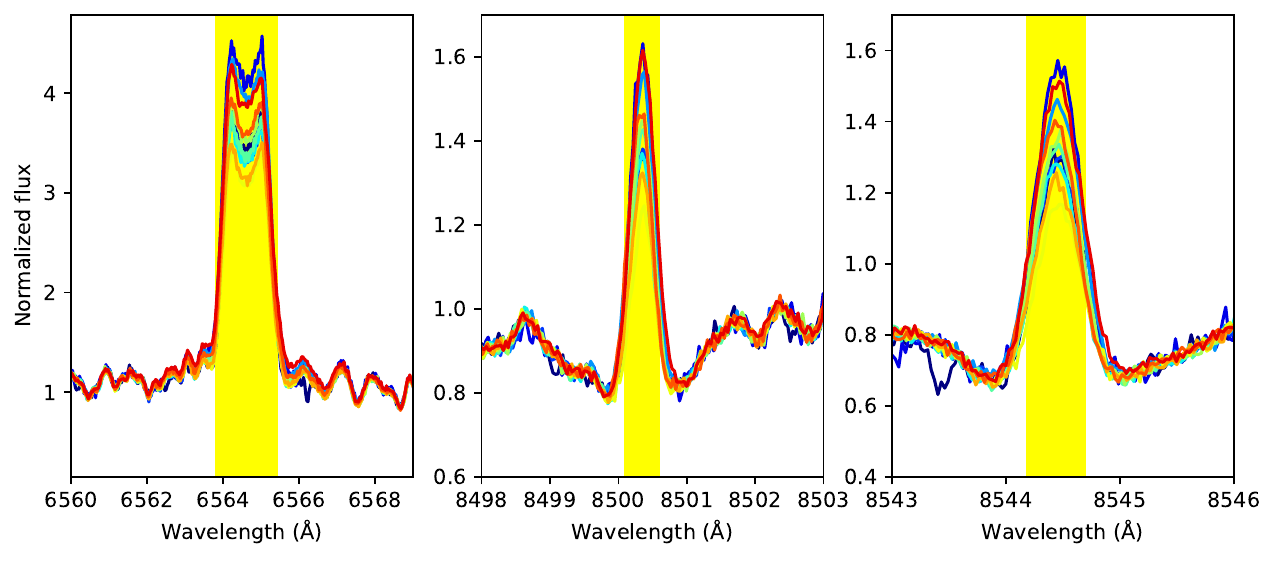}
\caption{Time domain observations of G\,80-21. The normalization of the spectra references the `Reference bands' from the Table~\ref{Table1}. The light yellow region indicates the fitting area, with its width based on the `Full width' from the Table~\ref{Table1}. The different colored spectral lines represent observations at different times, with the sequence from red to blue indicating an increase in time.}
\label{figure1}
\end{figure*}

The normalized time-domain spectra of the sample star are shown in Figure~\ref{figure1}. The different colored spectral lines represent observations at different times, with the sequence from red to blue indicating observations from early to late. The spectra are normalized using the `Reference bands' given in Table~\ref{Table1}. The light yellow region represents the fitting area, with its width determined by the `Full width' specified in Table~\ref{Table1}.

\begin{table}[htb]
\centering
\setlength{\tabcolsep}{5pt} 
\caption{ {\upshape Spectral line characteristics of H$_\alpha$ and \ion{Ca}{2}, as referenced from \cite{2019A&A...623A..24F}. The wavelengths provided represent vacuum wavelengths.}}
\label{Table1}

\begin{tabular}{ccccc}
\hline
\hline
Line & \multicolumn{2}{c}{Line band} & \multicolumn{2}{c}{Reference bands} \\
\cline{2-5}
     & Central & Full& Blue & Red \\
     & wave & width &  &  \\
     & (\AA) & (\AA) & (\AA) & (\AA) \\
\hline
H$_\alpha$   & 6564.62 & 1.6 & 6547.4--6557.9   & 6577.9--6586.4 \\
\ion{Ca}{2} 1      & 8500.35 & 0.5 & 8476.3--8486.3   & 8552.4--8554.4 \\
\ion{Ca}{2} 2      & 8544.44 & 0.5 & \multicolumn{2}{c}{Same as \ion{Ca}{2} 1 }\\
\hline
\end{tabular}
\end{table}

\section{Model construction}\label{sec:mod}

We start with the MARCS-OS model atmospheres, which consist of line-blanketed LTE model grids in either plane-parallel or spherical geometries \citep{2008A&A...486..951G}. The MARCS code is used to construct detailed models of the stellar photosphere, incorporating temperature, pressure, density, and chemical composition based on input parameters such as effective temperature, surface gravity, and metallicity. The model atmosphere of G\,80-21 is computed with parameters $T_{\rm eff}$ = 3473 $\pm$ 72\,K , log\,$g =$ 4.85 $\pm$ 0.18\,dex, and [Fe/H] = $-$0.18 $\pm$ 0.15\,dex. These stellar parameters are derived from \cite{2021A&A...656A.162M}, who determined atmospheric parameters for 348 M dwarf stars by combining optical and near-infrared spectroscopic data from the CARMENES spectrograph using spectral synthesis techniques and the BT-Settl model. The abundances of other elements are adopted from the scale of solar abundances, with values referenced from \citet{2009ARA&A..47..481A}.

Following the approach of \cite{1997A&A...326..287S} and \cite{2005A&A...439.1137F}, we extend the photospheric model by adding three sections of rising temperature to represent the lower and upper chromosphere and the transition region. To facilitate this extension, we increase the number of atmospheric layers in the model from 57 to 105. Our chromospheric model, similar to the PHOENIX model used in \cite{2019A&A...623A.136H}, is described by six free parameters: the column mass density at the onset of the lower chromosphere ($m_{\text{min}}$), which defines the location of the temperature minimum; the column mass densities and temperatures at the endpoints of the lower (${\rm log},m_{\text{mid}}$, $T_{\text{mid}}$) and upper chromosphere (${\rm log},m_{\text{max}}$, $T_{\text{max}}$); and the temperature gradient in the transition region ($\text{grad}_{\text{TR}}$). The LTE results for the chromosphere are calculated under hydrodynamic equilibrium. We then use the MULTI computer program of Scharmer and Carlsson \citep{1985JCoPh..59...56S,1985ASIC..152..189S} to perform the non-local thermodynamic equilibrium (NLTE) calculations for the whole atmosphere structures. The MULTI results combined with magnetic fields are used as the initial atmosphere conditions to RH1.5D code\citep{2015A&A...574A...3P} and the whole atmosphere structures are recalculated optimizing the depth scale based on optical depth, density, and temperature gradients. The RH1.5D model, like the PHOENIX model, allows for detailed spectral synthesis and accurate representation of stellar atmospheres.

The observations of the Sun reveal that the stellar surfaces exhibit both active and quiet regions \citep[e.g.][]{2024ApJ...966..197D}. Our method involves adjusting the ratio of active to quiet regions in the model spectra to agree with the observed spectra. This allows us to determine the variation of the areas for active and quiet regions along the line of sight over time for the sample star. In this study, We select the spectrum of an inactive star with similar stellar atmospheric parameters to serve as the model for the quiet region.

\subsection{Inactive model}

\begin{figure*}[htb]
\includegraphics[width=1.0\hsize]{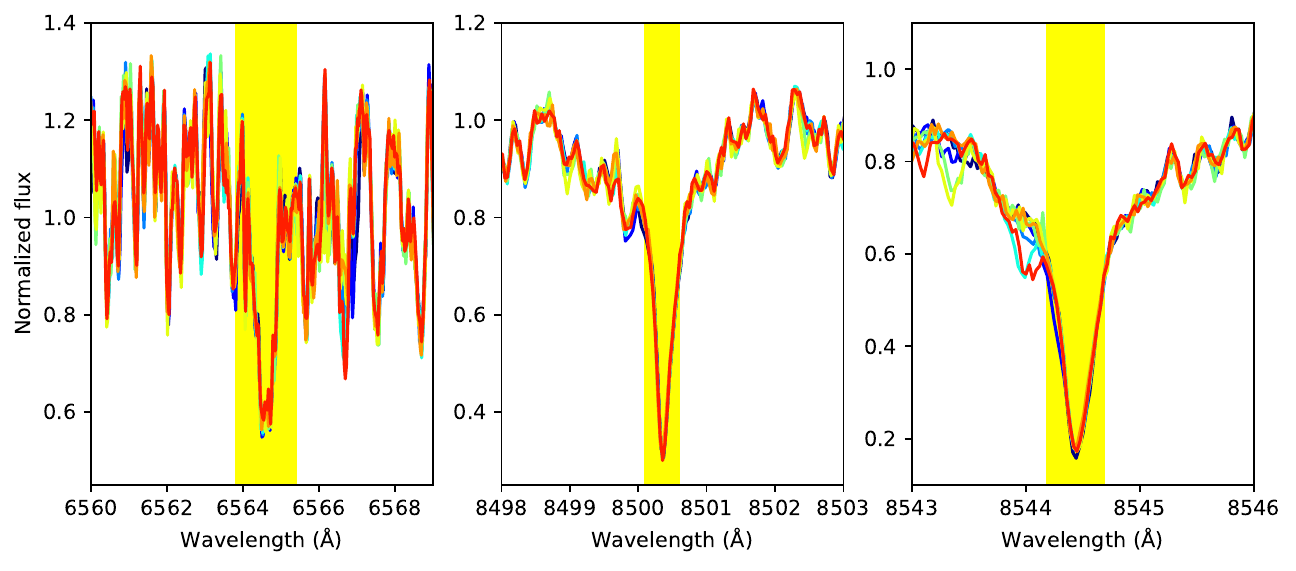}
\includegraphics[width=1.0\hsize]{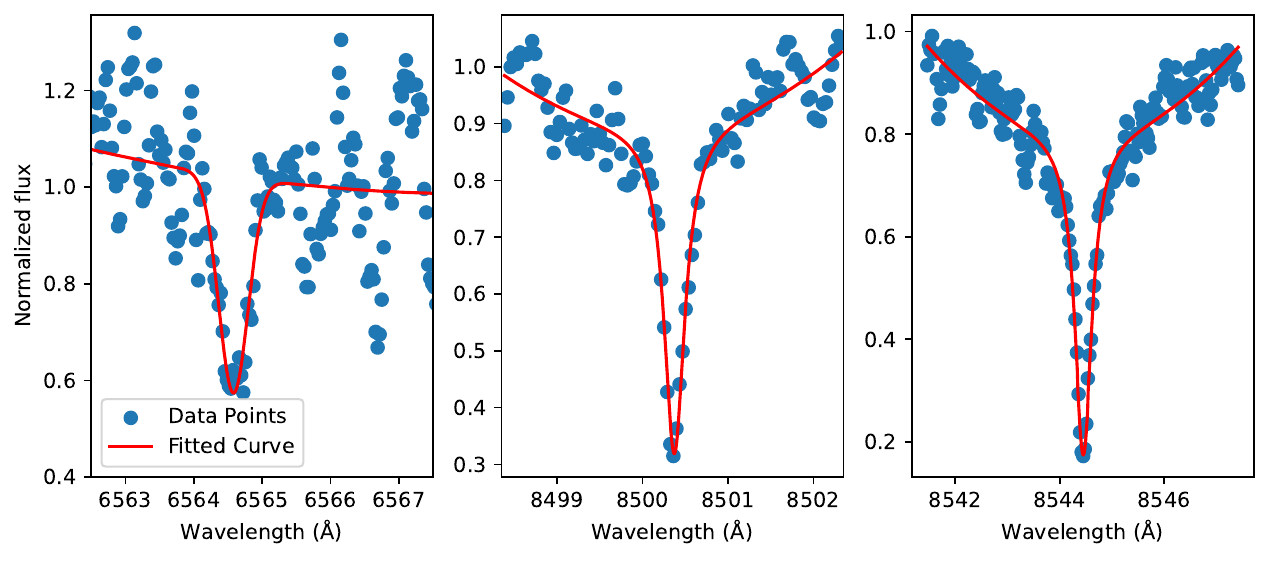}
\caption{The spectra of an inactive star of LP\,819-17. In the upper panel, the different line colors and the yellow-highlighted region have the same information as in Figure \ref{figure1}. Spectral analysis of LP\,819-17 showing consistent \ion{Ca}{2} IRT and H$\alpha$ lines without core emission (upper panels) and Lorentzian line profile fits (lower panels), serving as a quiet region model for star G,80-21 with SNR of 106 for H$\alpha$ and 117 for \ion{Ca}{2} IRT.Scatter points in the lower panel represent observed data with the highest signal-to-noise ratios (SNR), specifically SNR = 106 for H$_\alpha$ and 117 for \ion{Ca}{2} IRT lines. The red curves show Lorentzian line profile fits to these spectral lines.}
\label{figure2}
\end{figure*}

From the CARMENES catalog, we select the star LP\,819-17 as our model for an inactive star. According to the \cite{2021A&A...656A.162M}, this star has stellar atmospheric parameters of $\rm{T}_{eﬀ}$ = 3474 $\pm$ 23\,K , log\,$g =$ 4.85 $\pm$ 0.07\,dex, and [Fe/H] = $-$0.11 $\pm$ 0.07\,dex, which are very similar to those of our active sample star. By combining multiple observations (upper panels of Figure \ref{figure2}), we find that the spectra of this star in the \ion{Ca}{2} IRT and H$_\alpha$ lines are highly consistent, with no emission at the core of the \ion{Ca}{2} lines. Therefore, the spectrum of this star is well-suited to serve as the quiet region model for our sample star G\,80-21.

From multiple spectra of LP\,819-17, we select the spectrum with the highest SNR, specifically with an SNR of 106 and 117 for the H$_\alpha$ and \ion{Ca}{2} IRT lines, respectively. Using this spectrum, we apply a Lorentzian line profile to fit the \ion{Ca}{2} and H$_\alpha$ lines, as shown in lower panels of Figure~\ref{figure2}. The final resulting curve serves as our quiet region model.

\subsection{Active model set}

The chromospheric model is described by six free parameters. While the most straightforward approach to obtain a model grid that uniformly covers the typical ranges for all free parameters, even a moderate sampling of ten grid points per parameter results in a grid with ($10^6$) elements, leading to excessively high computational demands\citep{2013A&A...553A...6H}. The challenge lies in identifying reasonable parameter ranges and exploring these with multiple models. Initially, we calculated limiting cases, such as models that simultaneously show all spectral lines in absorption or emission, representing inactive or active chromospheric models. By visually comparing these models with the observed spectra, we identify the most promising parameter ranges for further exploration.

\section{Results and discussions}\label{sec:result}

\subsection{The impact of magnetic fields}
G\,80-21 is an M3.0-type star. \citet{2019A&A...626A..86S} analyzed the Ti and FeH spectral lines, and found that its surface magnetic field strength is approximately 3.2\,kG, with a projected rotational velocity of 5.7\,km$\,\rm{s^{-1}}$. The magnetic field measurements indicate that its distribution is relatively uniform, and there is a positive correlation between magnetic field strength and rotational velocity. This finding is consistent with the dynamo theory of stellar magnetic fields\citep{2007AcA....57..149K}.

\begin{figure*}[htb]
\begin{minipage}{0.9\textwidth}
    \includegraphics[width=\textwidth]{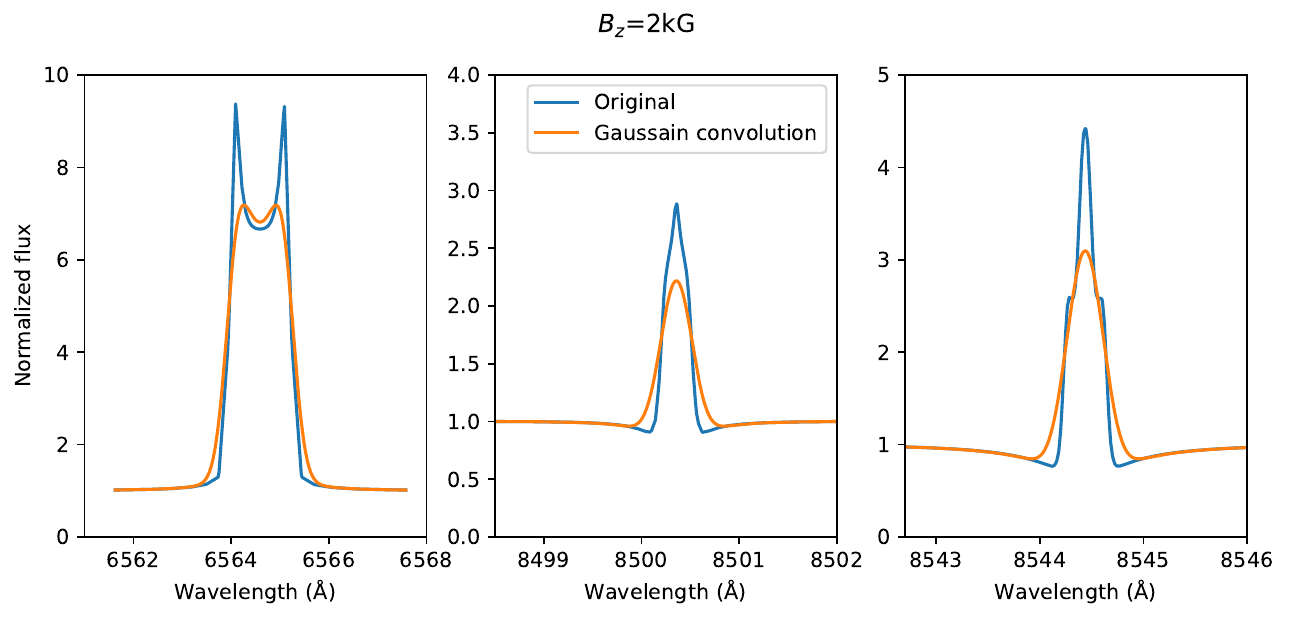}
\end{minipage}
\begin{minipage}{0.9\textwidth}
    \includegraphics[width=\textwidth]{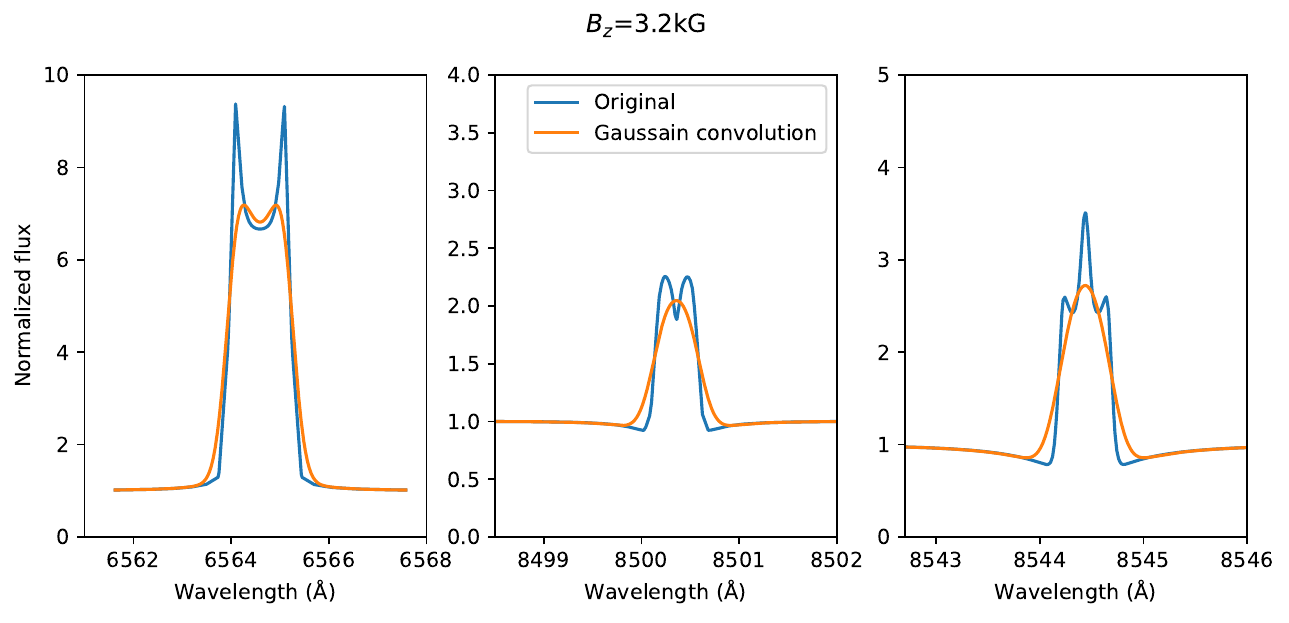}
\end{minipage}
\begin{minipage}{0.9\textwidth}
    \includegraphics[width=\textwidth]{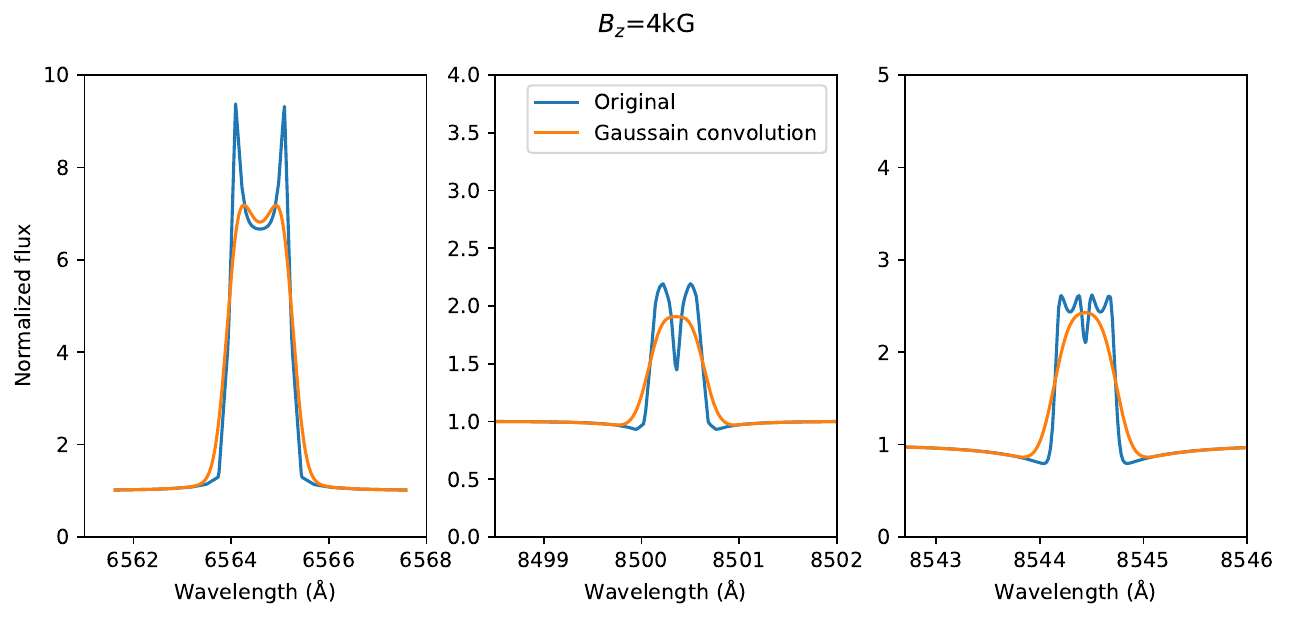}
\end{minipage}
\caption{The impact of magnetic fields of RH1.5D model. Theoretical spectral line profiles of H$\alpha$ (left column) and \ion{Ca}{2} IRT lines at 8498 \AA (middle column) and 8542 \AA (right column) under varying magnetic fields of 2 kG (top row), 3.2 kG (middle row), and 4 kG (bottom row). The blue curves represent original theoretical spectra without considering observational resolution effects (`Original'), while orange curves show these spectra convolved with observational resolution (`Observational convolution').}
\label{figure3}
\end{figure*}

In the RH1.5D model, the magnetic field can be included as input parameters. We select an active model structure to investigate the impact of magnetic fields on the chromospheric spectra of the \ion{Ca}{2} IRT and H$\alpha$ lines. As shown in Figure~\ref{figure3}, we adopt the average magnetic fields of 2, 3.2, and 4\,kG in the line-of-sight direction to this model structure. The theoretical spectral line profiles of H$\alpha$ (left column) and \ion{Ca}{2} IRT lines at 8498 \AA (middle column) and 8542 \AA (right column) are shown under varying magnetic fields of 2\,kG (top row), 3.2\,kG (middle row), and 4\,kG (bottom row). The blue curves represent original theoretical spectra without considering observational resolution effects (‘Original’), while orange curves show these spectra convolved with observational resolution (‘Observational convolution’). It can be seen that the line profile and intensity of H$\alpha$ remain unchanged regardless of magnetic field strength; however, significant Zeeman splitting is evident in both \ion{Ca}{2} IRT lines as they respond to increasing magnetic fields. When the theoretical spectra are convolved to the observational resolution, it is found that the \ion{Ca}{2} IRT lines appear as Gaussian profiles for magnetic fields of 2 and 3.2\,kG.

Based on the observed and simulated spectra of this sample star, the Zeeman effect induced by the magnetic field can only be detectable in the \ion{Ca}{2} IRT lines, affecting only the line profiles while not influencing the intensity. Additionally, due to instrumental and other line broadening factors, we cannot observe distinct double-peaked profiles for magnetic fields of 3.2\,kG or less.

\subsection{Models combination}
\subsubsection{One active region}

\begin{figure}[htb]
\includegraphics[width=1.0\hsize]{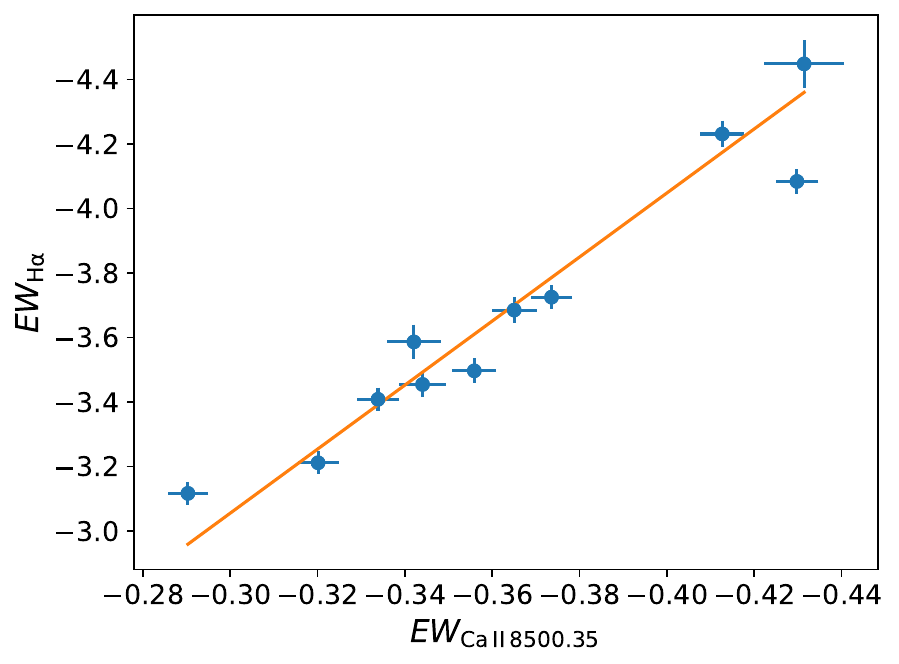}
\caption{The EW of \ion{Ca}{2} (8540\,\AA) line v.s. that of H$_\alpha$ line.}
\label{figure4}
\end{figure}

Figure~\ref{figure4} shows the distribution of the equivalent widths of H$_\alpha$ line versus the \ion{Ca}{2} (8500\,\AA) line for the active M dwarf star G\,80-21. It is evident from the figure that the emission intensity of \ion{Ca}{2} line increases linearly with the emission strength of H$_\alpha$ line, with a slope of 10, which indicates that the emission intensity of H$_\alpha$ line varies more significantly. Additionally, Figure~\ref{figure1} reveals a prominent double-peaked structure in the H$_\alpha$ emission line. This double-peaked structure is not unique for this star; we have also observed this situation in other active M dwarf stars observed by CARMENES, such as GJ\,1289 and AU Mic.

By organizing the RH1.5D theoretical model library, we find that the H$_\alpha$ double-peaked structure exists in models with a steep temperature rise in the lower chromosphere and a plateau in the upper chromosphere layers. However, when we attempt to fit the observed spectra by combining this model with the inactive model which mentioned in section~4.1, the resulting \ion{Ca}{2} line intensity is always much higher than that of the observed. This is because models with an H$_\alpha$ double-peaked structure often also exhibit very strong \ion{Ca}{2} lines, making it challenging to find a suitable proportion of active regions to fit the observed spectra accurately. Thus, an additional active region may be necessary to find an appropriate combination that fits the observed spectrum.

\subsubsection{Two active regions}

\begin{figure}[htb]
\includegraphics[width=1.0\hsize]{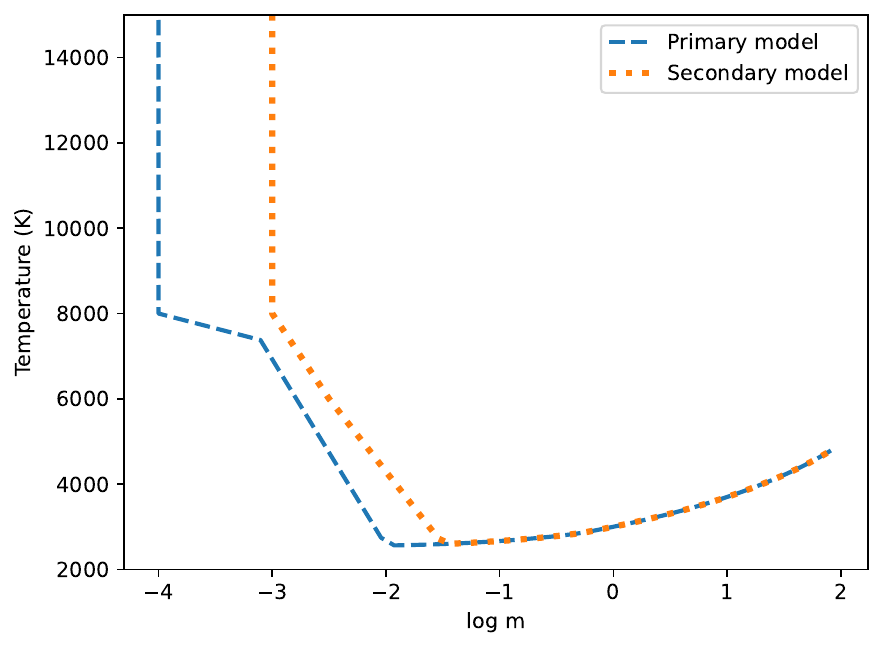}
\caption{Temperature structures of the primary and secondary models.}
\label{figure5}
\end{figure}

After thoroughly exploring our theoretical model library, we select the two most suitable active regions, which shown in Figure~\ref{figure5}. In Figure~\ref{figure5}, the blue and orange lines represent the density-temperature profiles of the two chromospheric activity structures, respectively. The structure represented by the orange line propagates less in the chromosphere, and exhibits stronger emission intensity. In contrast, the blue line corresponds to a structure similar to the solar VAL C model \citep{1981ApJS...45..635V}, characterized by a steep temperature rise in the lower chromosphere, and a plateau in the upper chromosphere, but with a larger $\log\,m_{\text{max}}$ in our model. As previously mentioned, the chromospheric spectrum generated by the blue structure shows a distinct double-peaked H$_\alpha$ line.

When these two active regions are combined with an inactive area, they provided a good fit for the H$_\alpha$ lines, and significantly improved the fitting for the \ion{Ca}{2} IRT lines compared to using a single active region model as discussed in the previous section. However, the emission intensity of the \ion{Ca}{2} IRT combination lines is still higher than that of the observed.

At this point, we notice that the photospheric Ca abundance used in the models is still set to the solar values, which might not be accurate. The [Fe/H] for this star is $-0.18\pm0.15$\,dex \citep{2021A&A...656A.162M}. According to APOGEE’s statistical results \citep{2022ApJ...927..123S}, the [Ca/Fe] ratio for M dwarfs generally ranges from $-$0.1 to 0.3\,dex within the metallicity range of $-0.4$ to 0.0\,dex. Given this information, the uncertainty range for [Ca/H] for this sample star is between $-0.4$ and 0.3 dex.

\begin{figure*}[htb]
\includegraphics[width=1.0\hsize]{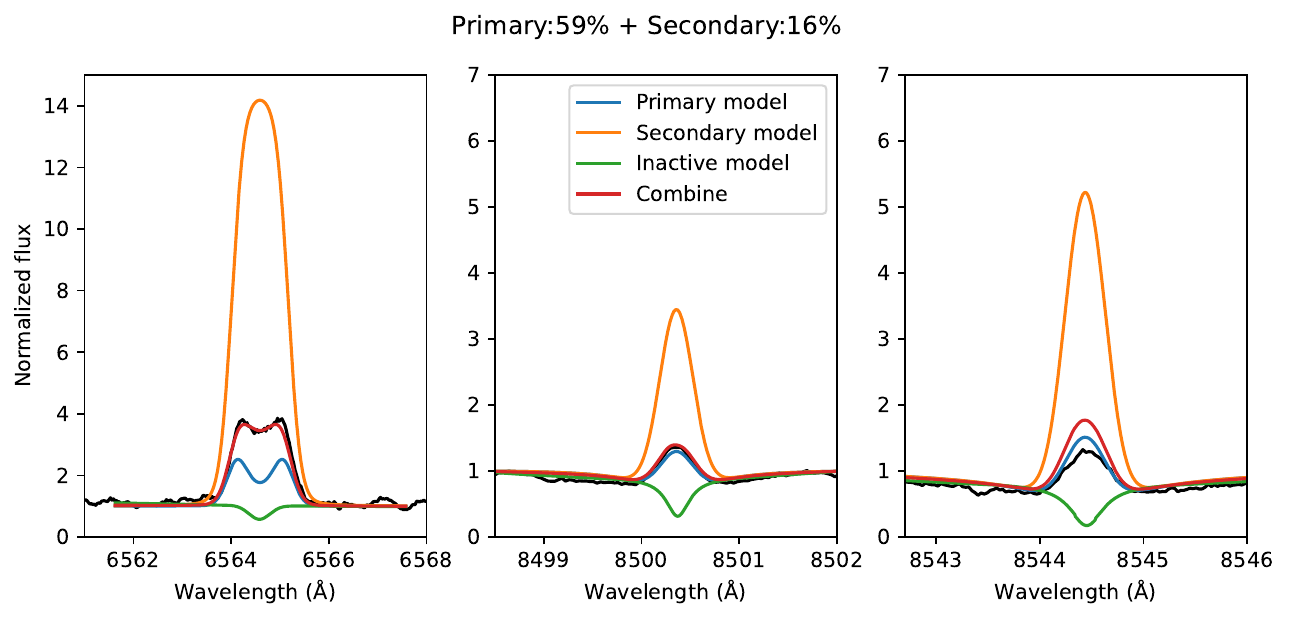}
\caption{The fitting results for the first observed spectrum of the sample star are shown. The black line represents the observed data from the first observation, while the blue and orange lines denote the primary and secondary active models, respectively. The green line represents the inactive model, and the red line indicates the combined spectrum.}
\label{figure6}
\end{figure*}

Figure~\ref{figure6} shows the fitting results of the first observed spectrum of the sample star. In the figure, the blue and orange lines represent the primary and secondary active regions, respectively; the green line represents contribution of the inactive region, and the red line represents the combined spectrum. It can be seen that when the photospheric Ca abundance is set to [Ca/H] $= -$0.4\, dex, the combined spectrum fits the observed spectrum (in black) well for both the blue \ion{Ca}{2} IRT line (8500\,\AA) and the H$_\alpha$ lines under a appropriate filling factor. However, the intensity of the red \ion{Ca}{2} IRT line at 8544\,\AA\ is still slightly higher than that of the observed one. Moreover, the ratios of different zones can be derived from Figure~\ref{figure6}. The primary active region, which corresponds to the weaker emission structure, accounts for 59\% of the total. The secondary active region, representing the stronger emission structure, comprises 16\%, while the inactive region makes up the remaining 25\%.
Additionally, we have tested the effect of parameter uncertainties on the final results. Specifically, we found that the uncertainty in $T_{\rm eff}$ (72\,K) affects the fractional area of the primary active region by 5\% and the secondary active region by 2\%. The uncertainty in log,$g$ (0.18\,dex) affects the primary active region by 13\% and the secondary active region by 3\%, while the uncertainty in [Fe/H] (0.15\,dex) affects the primary active region by 4\% and the secondary active region by 2\%.

\begin{figure}[htb]
\includegraphics[width=1.0\hsize]{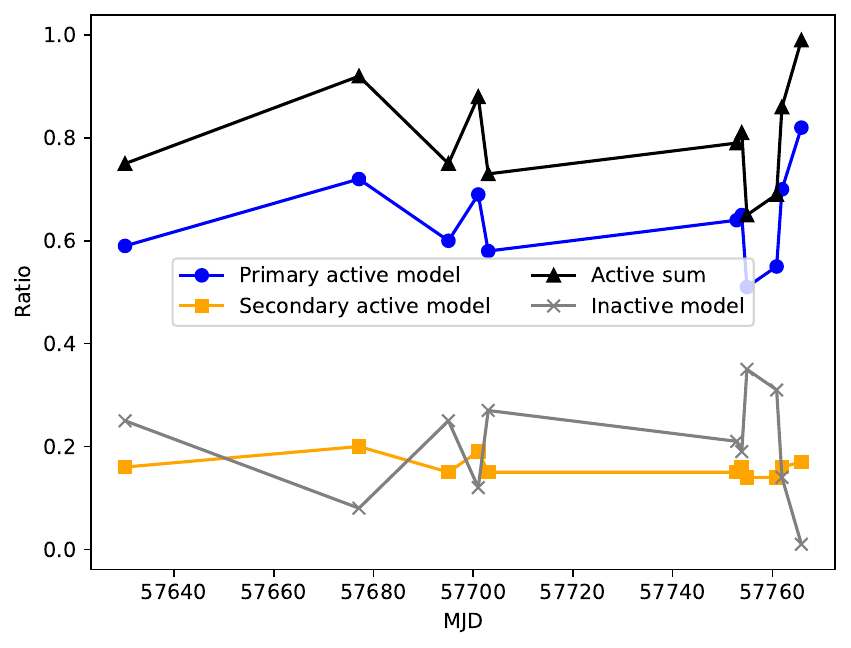}
\caption{The observed spectra at different times were fitted by adjusting the proportions of each model in the synthetic spectrum. }
\label{figure7}
\end{figure}

The observed spectra at different times can also be well fitted by adjusting the proportions of each region in the synthetic spectrum. Figure~\ref{figure7} illustrates the variation trends of the proportions of each region in different observed spectra. In the figure, the blue and orange points represent the primary and secondary active regions, respectively, while the black points indicate the combination of the two active regions. The gray points represent the inactive region. From the figure, it can be seen that the minor active component consistently comprises approximately 18\% of the total (ranging from 14\% to 20\%), while the major active component occupies a variable proportion, ranging from 51\% to 82\%. 
According to \citet{2018A&A...612A..49R}, the inclination angle of the sample star’s orbit is 55$\pm$16 degrees.
This suggests that the minor active component likely originates from the polar regions of the star, and does not move out of our line of sight during the star’s rotation. Therefore, its contribution to the spectrum remains relatively stable. This interpretation aligns with recent findings by \citet{2023A&A...680L..17K}, who observed magnetic field strengths of 3.0–3.5 kG near the poles, consistent with the 3.2 kG of our sample star. On the other hand, the major active component is unevenly distributed from the equator to the poles and may also vary with the star’s surface activity. This leads to a larger variation in its contribution to the spectra. 



\section{Conclusions}\label{sec:con}
Based on the observed high-resolution spectra from the CARMENES project we investigate the active regions of the M3.0V star G\,80-21 by utilizing the synthetic spectra produced with the RH1.5D radiative transfer code. The successful application of the RH1.5D code demonstrates its capability in accurately synthesizing the chromospheric spectrum. The conclusions are as follows:
\begin{itemize}
\item[$\bullet$] The optimal fit for the observed spectrum is achieved by employing two active regions in conjunction with an inactive region, where the [Ca/H] = $-$0.4 is adopted.
\item[$\bullet$] The combination is capable of fitting all the observed data across varying ratios.
\item[$\bullet$] The minor active component consistently comprises approximately 18\% of the total(14\%$\sim$20\%), while the major active component occupies a variable proportion, ranging from 51\% to 82\%.
\end{itemize}
In the future, we will apply this method to study other types of active stars, which will help us understand the physical mechanisms driving the stellar activity.

\section*{Acknowledgements}
This research is supported by the National Natural Science Foundation of China under Grant Nos. 12090040/4, 12373036, 12022304, 11973052, 12173058, the National Key R$\rm\& $D Program of China No.2019YFA0405502, the Scientific Instrument Developing Project of the Chinese Academy of Sciences, Grant No.\,ZDKYYQ20220009, and the International Partnership Program of the Chinese Academy of Sciences, Grant No.\,178GJHZ2022047GC.

\bibliography{sample631}{}
\bibliographystyle{aasjournal}



\end{document}